# PeppyChains: Simplifying the assembly of 3D-printed generic protein models

Promita Chakraborty (QuezyLab)

Peppytides is a coarse-grained, accurate, physical model of the polypeptide chain [1-4]. I have shared instructions to make your own polypeptide chain and STL files of Peppytides in MAKE last year (Jan 2014 issue). However, Peppytides involves a lot of steps and assembly of units. People were asking me, "Is there any easier way to make these models?". I have been at several workshops, hands-on sessions, talks and a science events with Peppytides. From the feedback that I got everywhere, most of the makers, teachers and students want something that is 3D-Print-&-Go, or at least easier to make, even at the cost of some features.

Because of these requests, I have designed a new version of the model, named **PeppyChains**, in which the backbone chain of the model can be 3d-printed as a single unit. PeppyChains design eliminates the assembling of parts to form the backbone chain, but with the cost of losing the ability to use bias-magnets to favor certain phi/psi angles in the backbone. I have been demonstrating PeppyChains at various talks and workshops along with Peppytides, and I have received requests for making the STL file public. On popular demand, here I am sharing this STL and providing the step-by-step instructions to make PeppyChains. Just add the hydrogen-bond magnets, paint and color-code the atoms, and you are ready to go! No drilling. No time consuming assembly. Peppytides and PeppyChains are now available for purchase at www.quezylab.com, but if you want to make one by yourself, you are welcome.

Here is a picture of a class interacting with Peppytides and PeppyChains during my talk followed by a hands-on workshop at Foothill College, Los Altos, summer Nanocamp 2014.

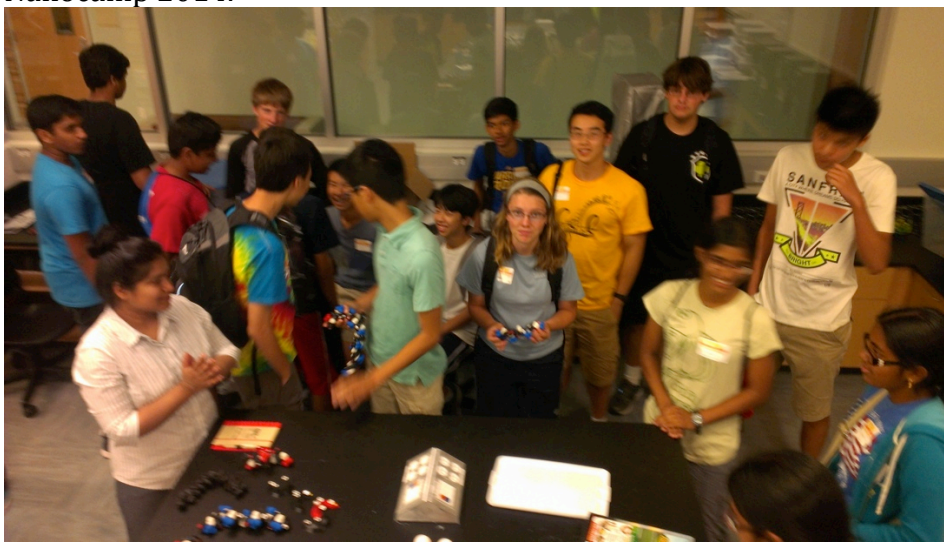

Another strong feature I have added in PeppyChains is to make detachable side chains. Thus, anyone would be able to interchange side chain units to make and fold different sequences of amino acids on-the-fly. This feature is very useful if someone wants to make a specific sequence to experiment with protein folding, since a particular sequence represents a unique 3D structure and hence a specific protein. The previous version, Peppytides, had only one specific type of amino acid, Alanine, all along the chain. Once Peppytides was assembled, there was no option to remove and replace side chains with other types. **PeppyChains provides a generic platform to make different protein.** Side chains are interchangeable so PeppyChains can be used to work with all types of amino acids[1]. I included STL files for 3 types of amino acids: Alanine, Glycine and Valine. For future, I am planning to provide STL files for the rest of the amino acid side chains as well.

This version of PeppyChains is a 7-mer, which means that we would be able to fold it into an alpha-helix of 2 turns (@ 3.6 amino acids per turn). You can add multiple models, one on top of another to make bigger helices.

By 3D-printing one single piece of backbone chain, we are losing one feature: that is biasing the backbone chain with magnets to favor certain angles for folding only into alpha-helices or beta-sheets. In previous Peppytides, we add magnets and then assemble the units to form the backbone chain. Since backbone in PeppyChains is printed as a whole at once, it is not possible to put bias-magnets now. I will be adding this feature in a future version of PeppyChains. Please note that hydrogen bond (H-bond) magnets are still used in PeppyChains, similar to Peppytides.

In PeppyChains, amide and alpha carbon units are connected by rotatable alternate phi and psi bonds. Each component in the backbone chain are connected together in such a way that they can rotate freely. Therefore, soluble support material is crucial for 3D printing. Please see Figure 1 for more details.

---

[1] The interchangeable side chain feature has been added in **Peppytide 2 (Peppytide with interchangeable side chains)**. While Peppytides has established that foldable physical model of proteins can be built and folded with accuracy, it was only a model of poly-alanine. Peppytides 2 extends it and enables interchangeable side chains. Stay tuned for this new features and updated STL files. Please note that Peppytides 2 has the rotational bias, absent in PeppyChains but it still has the almost same assembly process as in Peppytides.

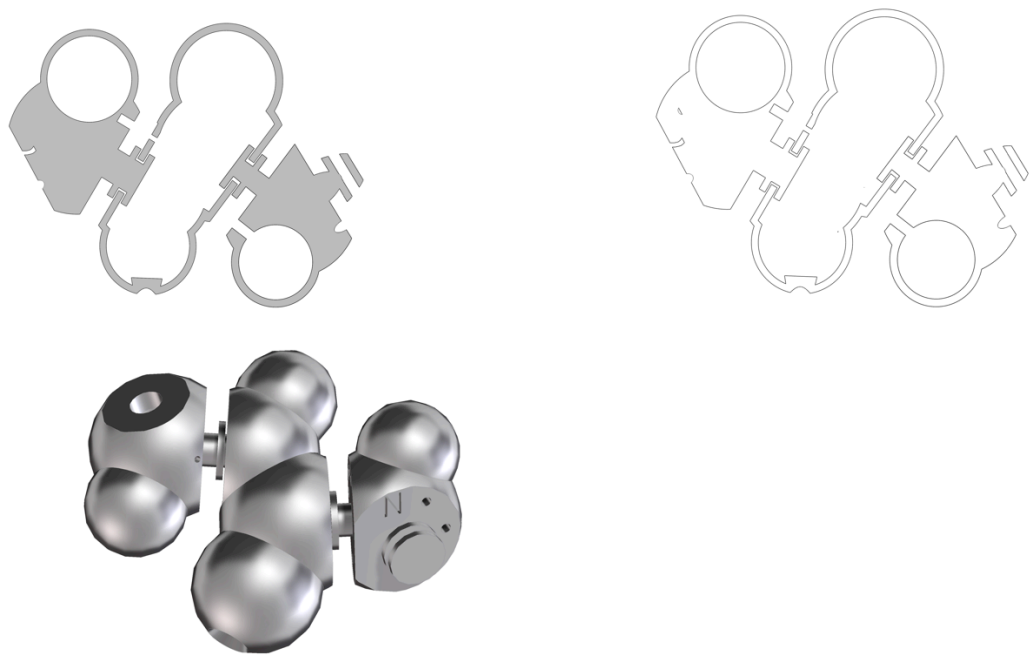

Figure 1: PeppyChain backbone units: (top) cross-section showing design of joints for rotation in repeating pattern of alphaCarbon-amide-alphaCarbon; (bottom) 3D CAD drawing

PeppyChains 3D design enables us to print the entire chain as a whole using a single STL file. After printing, it is a single piece forming the flexible backbone without any manual assembly (such as using bolts and screws to connect amide and alpha carbon units in Peppytides).

The STL files for the 7-mer backbone and the 3 side chains are provided below.

**Tools and materials:**
- **M1:** I tested with uPrint Plus (Black and blue/yellow materials)
- **M2:** Acrylic paints (Red for oxygen, Blue for nitrogen, white for hydrogen)
- **M3:** Magnets for H-bond [3/16″ D x 1/8″ H; Supplier: K&J Magnetics http://www.kjmagnetics.com/; Part number: D32-N52]; Quantity: 19 (16 for the 8 amides in backbone and 3 in helix-template).
- **M4:** Glue that holds metal and plastic together (Epoxy JB-weld)
- **M5:** Sandpaper (220 grit) for sanding magnets for a better grip for the glue (WARNING: Do not miss this step, otherwise your magnets might come out after a few uses) **Tips:** Using a rough concrete slab might work better.
- **M6**: backbone STL (file: PC1_chain)
- **M7:** – side chain STL (files: PC2_alanine, PC3_glycine, PC4_valine)
- **M8:** – Helix template STL (file: PC5_helixTemplate)

**Steps Summary:**
        **Step 1-3:**    3D printing parts;
        **Step 4:**       Color/paint atoms;
        **Step 5 & 6:**  Sand & glue magnets to parts;

**Step 1: 3D print the backbone.**

Use a 3D printer that supports soluble support materials that can be melted away after 3D printing. I have used uPrint Plus and black color material.

In uPrint Plus, printing M6 took about 11 hours, and then a few hours of soaking and drying.

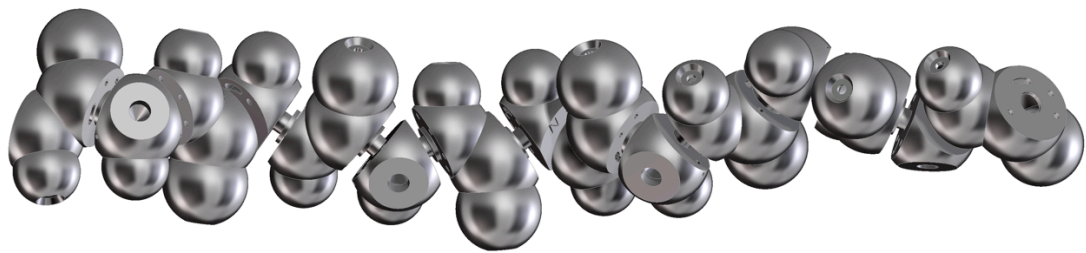

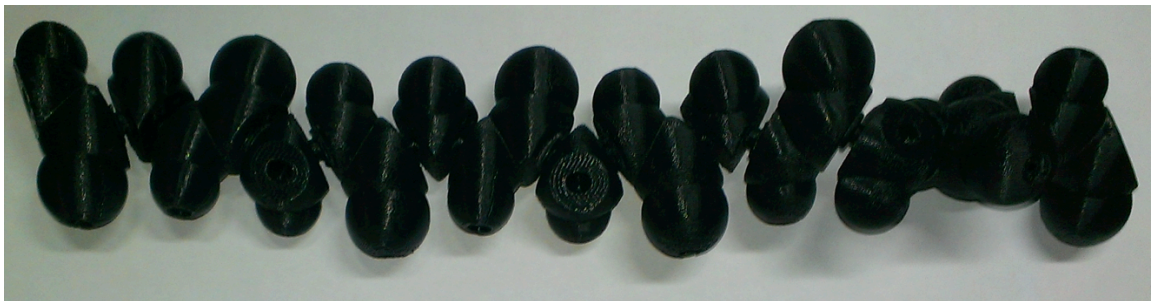

Figure 1: (top) CAD drawing for PeppyChains; (bottom) PeppyChain backbone 3D printed.

**Step 2: Printing the side chains**

Side chain units Alanine, Glycine, Valine (Figure 2, STL files: M7) are 3D-printed in a color of your choice so that they can be plugged into the backbone chain (blue here) – **preferably yellow** as these are all hydrophobic amino acids. Each piece takes about 30-40 minutes on an average for printing, then a few hours for soaking and drying.

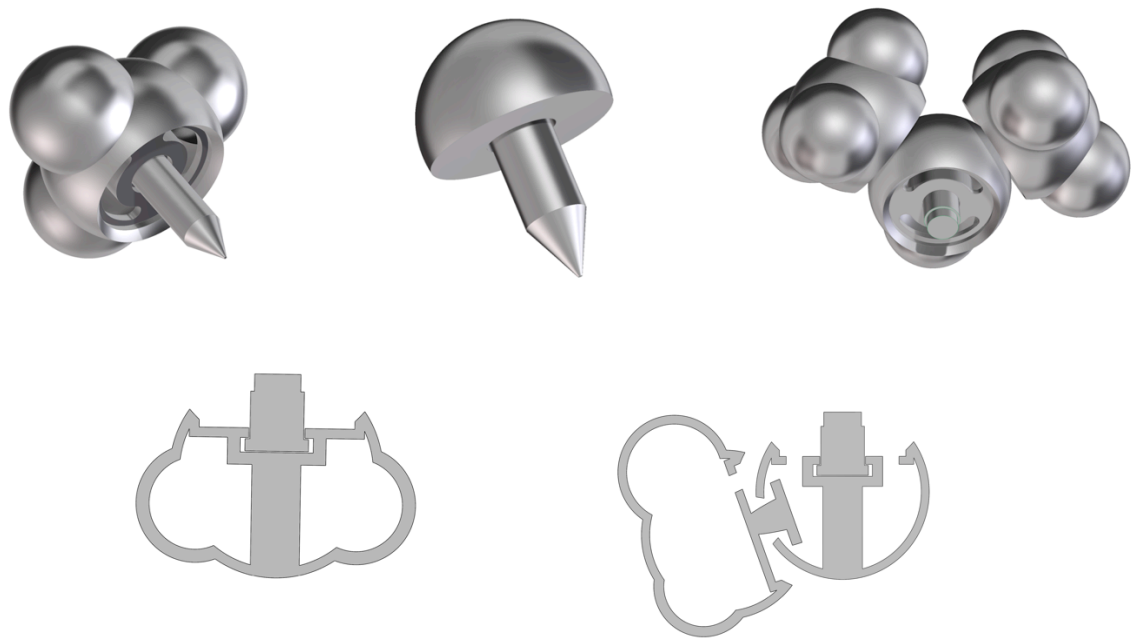

Figure 3: (top) Interchangeable side chains (Alanine, Glycine, Valine); (bottom) CAD drawing cross section (Alanine, Valine)

**Step 3: 3D print helix template**

This part is needed when you are experimenting and folding with PeppyChains (STL file: M8). For finishing touch (adding magnets), go to Step 6.

**Step 4: Paint to indicate elements oxygen, hydrogen, nitrogen atoms**

Use acrylic paints to indicate the different elements in the backbone amide units. (colors: Red for oxygen, Blue for nitrogen, white for hydrogen). This is the default color scheme used by the research community and most visualization tools.

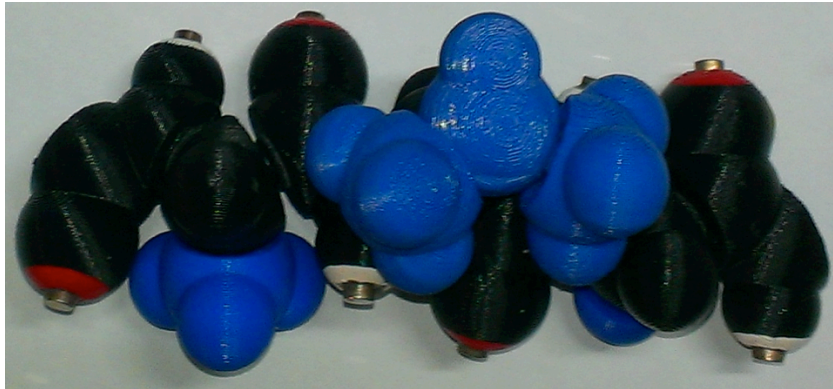

Figure 4: PeppyChains with replaceable side chains Alanine, Valine and Glycine in sequence

**Step 5:  Adding H-bond magnets in backbone**

Sand the bottom face of the H-bond magnets M3 (3/16" x 1/8") with 220 grit sandpaper to roughen the surfaces for effective adhesion. Next, glue the magnets onto the amides using M4 Epoxy (JB-weld), such that the oxygen atom (O) has the North pole up, and the hydrogen atom (H) has the South pole up. Leave for 24 hours for setting and drying.

**Step 6: Adding H-bond magnets to the helix template.**

• Sand three H-bond magnets M3 (3/16" x 1/8") and glue them onto the helix template using M4 Epoxy (JB-weld) with South poles up. These represent the hydrogen atoms (H) that will initiate and stabilize the folds in PeppyChains. Leave for 24 hours for setting and drying.

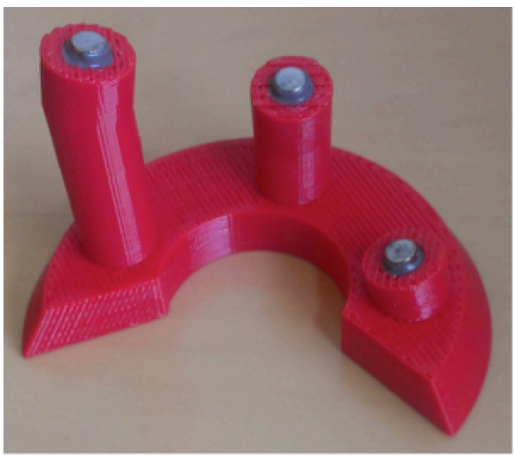

**Final discussions:**

The 7-mer PeppyChain is now ready for experimenting with by making different sequences of amino acids and studying how that guides folding. Go to the [Peppytides website](#) for directions on how to fold into the alpha-helices and beta sheets - you would now need the helix stand from step #6 to help in alpha-helix folding.

Please share your experience in folding and what other features you would like.